\def\year{2018}\relax
\documentclass[letterpaper]{article} %DO NOT CHANGE THIS
\usepackage{aaai18}  %Required
\usepackage{times}  %Required
\usepackage{helvet}  %Required
\usepackage{courier}  %Required
\usepackage{url}  %Required
\usepackage{graphicx}  %Required
\usepackage{amsmath}
\usepackage{natbib}
\begin{document}
\title{Do Political Detachment Users Receive Various Political Information on Social Media?}
\author{
Mitsuo Yoshida \\
Toyohashi University of Technology \\
Toyohashi, Aichi, Japan \\
\url{yoshida@cs.tut.ac.jp}
\And
Fujio Toriumi \\
The University of Tokyo \\
Bunkyo-ku, Tokyo, Japan \\
\url{tori@sys.t.u-tokyo.ac.jp}
}
\maketitle
\begin{abstract}
In the election, political parties communicate political information to people through social media.
The followers receive the information, but can users who are not followers, political detachment users, receive the information?
We focus on political detachment users who do not follow any political parties, and tackle the following research question:
do political detachment users receive various political information during the election period?
The results indicate that the answer is ``No''.
%Political detachment users did not receive various political information and could only receive information of very few political parties, as the followees only retweeted tweets of certain political parties on Twitter.
%Even users who do not actively obtain political party information were selectively obtaining information of political parties practically.
We determined that the political detachment users only receive the information of a few political parties.
\end{abstract}

\section{Introduction}

The partial amendment of the public officers election act of Japan in 2013 authorized election campaigning using the Internet.
In the latest election, the 48th general election for the Lower House in Japan 2017, social media, especially Twitter, was actively used.
The main feature of this election is that a new party named The Constitutional Democratic Party of Japan (@CDP2017) attracted numerous followers on social media\footnote{Bloomberg (2017): A 3-Day-Old Japanese Political Party Has Already Overtaken Abe’s on Twitter.}.
%Generally, if we get many followers, we can deliver information to many people.
The followers receive the information, but can users who are not followers, political detachment users, receive the information?
We are interested in how much political information has been received by political detachment users on social media.

Previous studies show that users are divided about access to political information (e.g.,~\cite{Adamic2005,Hayat2017,Hyun2014,Iyengar2009}).
Such studies mainly cover political information written in the news on social media.
In fact, its do not cover direct information from political parties.
Previous studies also show the relationship between the division of information and ideology (e.g.,~\cite{Barbera2015,Batorski2018,Dahlgren2005,Williams2015}).
In other words, the studies were analyzing users who tended to support particular political parties.
We focus on political detachment users who are not such users, and direct information from political parties.
In this paper, we use retweets and social graph on Twitter to address the following research question:
do political detachment users receive various political information during the election period?
%The results indicate that the answer is ``No''.
%We determined that the political detachment users only receive the information of a few political parties.
We discuss this phenomenon by using the information entropy.

\section{Data and Methodology}

\subsection{Data}

In this study, we use Japanese retweets on Twitter collected from 28 September\footnote{The Lower House in Japan was dissolved on this day.} to 23 October 2017.
This period encompasses the 48th general election for the Lower House in Japan.
These data were collected using the Twitter Search API\footnote{We constantly searched by query ``RT lang:ja''.} and consists of 42,651,648 retweets.

In order to focus only on data related to politics, we focus on the official accounts of political parties and selected the accounts of the six major parties.
%The parties are The Liberal Democratic Party of Japan (@jimin\_koho), The Constitutional Democratic Party of Japan (@CDP2017), The Party of Hope (@kibounotou), Komeito (@komei\_koho), The Japanese Communist Party (@jcp\_cc) and The Japan Innovation Party (@osaka\_ishin).
The accounts are @jimin\_koho, @CDP2017, @kibounotou, @komei\_koho, @jcp\_cc and @osaka\_ishin.
In the collected retweets, we only use 732,861 retweets (84,043 users) in which the tweets of these political parties have been retweeted.

We also use the social graph on Twitter.
First, we gathered users who are following the major parties.
Then, we created a set of users that combines these users and the users who have retweeted tweets of the major parties.
This set includes 460,683 users.
Finally, we gathered users who are following the 460,683 users on 10 November 2017 and build the social graph.
As a result, the social graph consists of 16,742,073 nodes (users) and 409,741,963 edges (followee-follower paths).

\subsection{Methodology}

We consider ``political detachment users'' as users satisfying the following:
they are not following any political party and have not retweeted any tweets of political parties.
There are 13,174,064 users~(78.7\%) as political detachment users, of the 16,742,073 gathered users.
The number of times a user received political party information is calculated as follows:
number of times that followees (followed users) of the user retweeted tweets of political parties.
We assume that the user has ``received'' political party information because it appears on the timeline of the user if the followees are retweeting tweets of political parties.

We use the information entropy to measure how much various political information political detachment users receive.
The information entropy $E$ is defined as follows:
\begin{align*}
E = - \sum_{p \in P} q_p \ln q_p
\qquad
q_p = \frac{C_p}{C}
\qquad
C = \sum_{p \in P} C_{q}
\end{align*}
where $P$ is a set of the major parties,
$C_p$ is the number of times a user received political information of the party $p$ ($\in P$).
The information entropy decreases as the number of times a user received political information becomes more biased.
In this paper, $E=0$ if a user only received information from a party, and $E=1.79$ if a user received information equally from all major parties.

\section{Results and Discussion}

\begin{figure}[t]
  \centering
  \includegraphics[width=0.99\linewidth]{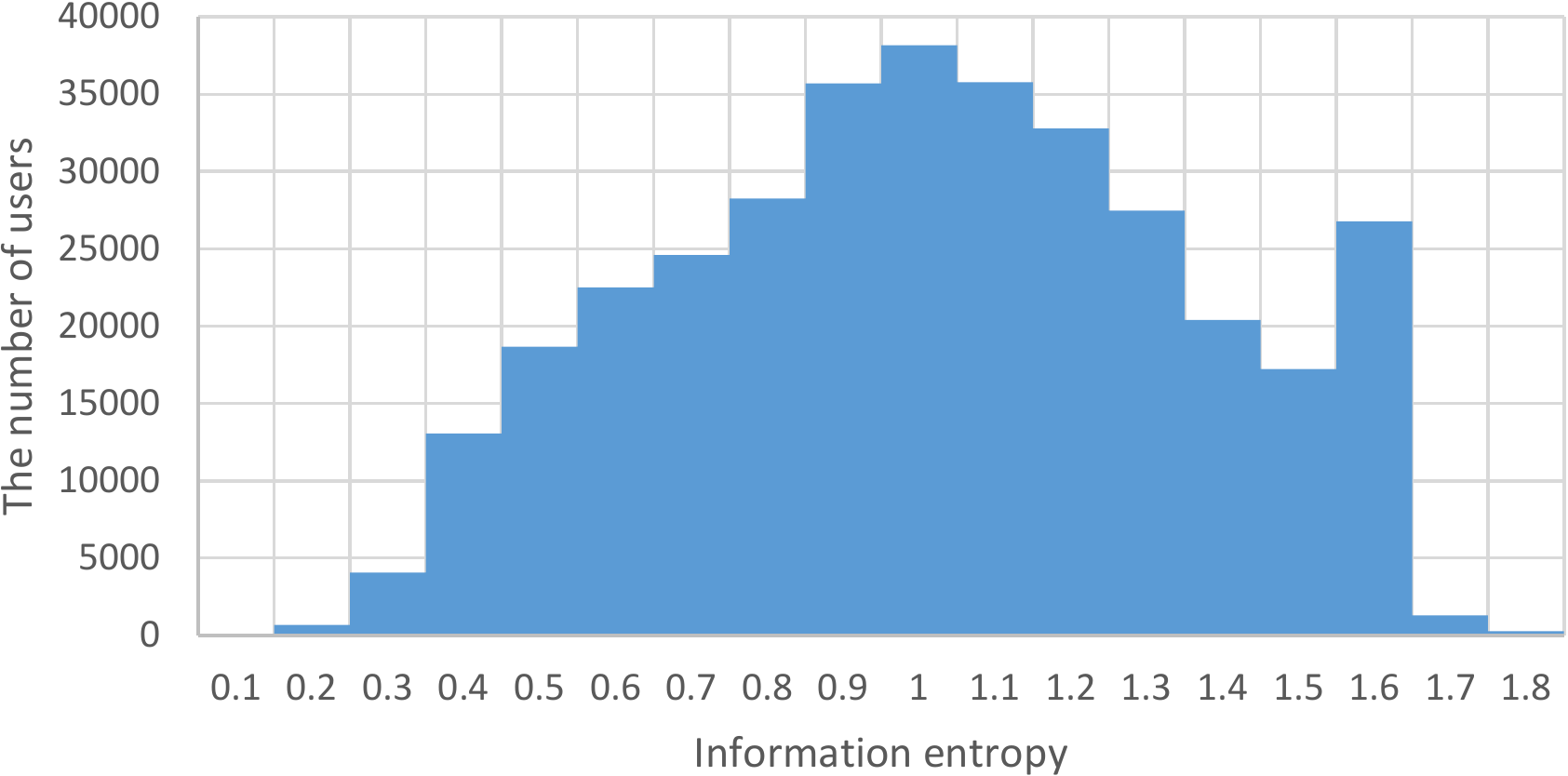}
  \caption{Distribution of information entropy ($0.1 \leq E$): the x-axis indicates the bin of information entropy, e.g., $0.1$ means $0.1 \leq E < 0.2$.}
  \label{fig:information-entropy}
\end{figure}

We computed the information entropy of each political detachment user satisfied $C > 0$.
Users satisfied $0 \leq E < 0.1$ were 12,826,532 users~(97.4\%) and $0.1 \leq E$ were 347,532 users~(2.6\%).
This means that most political detachment users do not receive various political information.
Figure~\ref{fig:information-entropy} shows the distribution of information entropy under $0.1 \leq E$.
Since there is a peak in $1.6$, there are also a few users who received political information almost equally.

\begin{figure}[t]
  \centering
  \includegraphics[width=0.99\linewidth]{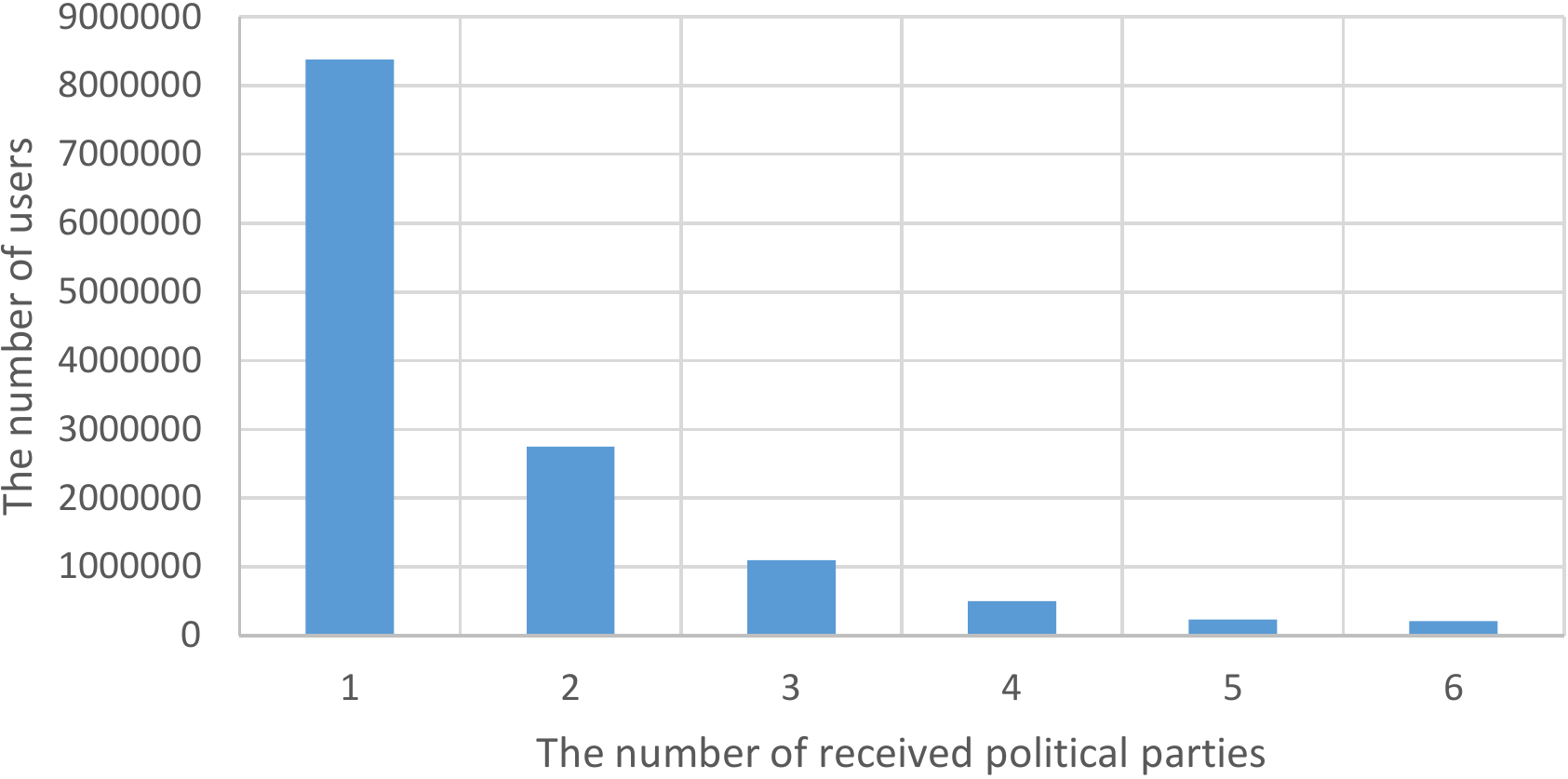}
  \caption{Distribution of the number of received political parties: many political detachment users only received information (tweets) of one party.}
  \label{fig:times-received}
\end{figure}

Figure~\ref{fig:times-received} shows the number of political parties from which a political detachment user received information.
8,381,477 political detachment users~(63.6\%) only received information (tweets) of one party.
Users who received all major parties total only 211,060 users~(1.6\%).

Political detachment users do not follow any political parties and cannot receive political information directly.
On the other hand, we analyzed the social graph and found that all political detachment users can receive information on which any political party through the followees.
However, political detachment users do not receive various political information and can only receive information of very few political parties, as the followees only retweet tweets of certain political parties.
Even users who do not actively obtain political party information are selectively obtaining information of political parties practically.

\section{Conclusion}

We focused on political detachment users who do not follow any political parties, and tackled the following research question:
do political detachment users receive various political information during the election period?
The results indicate that the answer is ``No''.
Political detachment users did not receive various political information and could only receive information of very few political parties, as the followees only retweeted tweets of certain political parties on Twitter.
Even users who do not actively obtain political party information were selectively obtaining information of political parties practically.

\bibliographystyle{aaai}
\bibliography{icwsm-w15}

\end{document}